\documentclass[runningheads]{llncs}

\usepackage{graphicx}

\usepackage{lipsum}

\usepackage{amsmath}
\usepackage{amssymb}
\usepackage{makecell}
\usepackage{multirow}
\usepackage{booktabs}
\usepackage{bbding}
\usepackage[marginal]{footmisc}

\usepackage{subfigure}
\usepackage{cite}
\usepackage{float}
\usepackage{color}

\begin{document}

\title{FundusGAN: A Hierarchical Feature-Aware Generative Framework for High-Fidelity Fundus Image Generation}
\titlerunning{FundusGAN}


\author{Qingshan Hou\inst{1,2} \and Meng Wang\inst{4} \and Peng Cao\inst{1,2}  \and Zou Ke\inst{4} \and Xiaoli Liu \inst{1,2} \and Huazhu Fu\inst{5} \and Osmar R. Zaiane\inst{3} }
\authorrunning{Q. HOU et al.}

\institute{Computer Science and Engineering, Northeastern University, Shenyang, China \and
Key Laboratory of Intelligent Computing in Medical Image of Ministry of Education, Northeastern University, Shenyang, China
\and Alberta Machine Intelligence Institute, University of Alberta, Edmonton, Canada\\
\and  Ophthalmology, Yong Loo Lin School of Medicine, National University of Singapore, Singapore\\
\and Institute of High Performance Computing, Agency for Science, Technology and Research
}

\maketitle

\begin{abstract}
Recent advancements in ophthalmology foundation models such as RetFound have demonstrated remarkable diagnostic capabilities but require massive datasets for effective pre-training, creating significant barriers for development and deployment. To address this critical challenge, we propose FundusGAN, a novel hierarchical feature-aware generative framework specifically designed for high-fidelity fundus image synthesis. Our approach leverages a Feature Pyramid Network within its encoder to comprehensively extract multi-scale information, capturing both large anatomical structures and subtle pathological features. The framework incorporates a modified StyleGAN-based generator with dilated convolutions and strategic upsampling adjustments to preserve critical retinal structures while enhancing pathological detail representation. Comprehensive evaluations on the DDR, DRIVE, and IDRiD datasets demonstrate that FundusGAN consistently outperforms state-of-the-art methods across multiple metrics (SSIM: 0.8863, FID: 54.2, KID: 0.0436 on DDR). Furthermore, disease classification experiments reveal that augmenting training data with FundusGAN-generated images significantly improves diagnostic accuracy across multiple CNN architectures (up to 6.49\% improvement with ResNet50). These results establish FundusGAN as a valuable foundation model component that effectively addresses data scarcity challenges in ophthalmological AI research, enabling more robust and generalizable diagnostic systems while reducing dependency on large-scale clinical data collection.

\keywords{Color fundus images \and Generative adversarial networks \and Disease diagnosis \and Foundation model.}
\end{abstract}

\section{Introduction}

Fundus image analysis is essential for the early detection and management of various ocular diseases such as diabetic retinopathy, glaucoma, and age-related macular degeneration~\cite{spaide2015retinal,ting2017development}. The development of foundation models in ophthalmology, such as RetFound~\cite{zhou2023foundation}, has demonstrated significant potential for improving diagnostic accuracy and efficiency in clinical settings. However, these advanced models face a critical challenge: they require massive amounts of high-quality annotated fundus images for effective pre-training~\cite{li2021applications,abramoff2018pivotal}. In real-world scenarios, acquiring large-scale, diverse, and well-annotated fundus image datasets is extremely challenging due to privacy concerns, annotation costs, and the inherent scarcity of certain pathological cases~\cite{natu2021privacy,kadambi2021achieving}.

Generative approaches have emerged as promising solutions to address data scarcity issues in medical imaging~\cite{ibrahim2025generative,costa2017end}. Current methods for fundus image synthesis, however, struggle with several key limitations. First, existing generative models often fail to capture the hierarchical nature of fundus structures, which span multiple scales from large anatomical features (optic disc, macula) to fine-grained pathological details (microaneurysms, hemorrhages)~\cite{zhao2015retinal,fu2018disc}. Second, most approaches lack the ability to simultaneously preserve global anatomical coherence while accurately representing subtle disease-specific lesions~\cite{guibas2017synthetic,saeed2021accuracy}. Third, existing methods frequently produce artifacts and unrealistic textures that limit their clinical utility for training diagnostic models~\cite{zhao2018synthesizing}.

To address these challenges, we propose FundusGAN, a hierarchical feature-aware generative framework specifically designed for high-fidelity fundus image synthesis. Our approach makes three key innovations: (1) We incorporate a Feature Pyramid Network (FPN) in the encoder to comprehensively extract and fuse multi-scale information, enabling the model to represent hierarchical features ranging from large-scale anatomical structures to subtle lesions~\cite{lin2017feature}. (2) We introduce a latent content vector mapping mechanism that adaptively decomposes and reconstructs different structural elements of fundus images, ensuring both anatomical integrity and pathological detail preservation. (3) We optimize the generator architecture with dilated convolutions and strategic adjustments to upsampling operations, significantly enhancing the model's ability to generate medically relevant structural information.

Extensive experiments on multiple benchmark datasets (DDR, DRIVE, and IDRiD) demonstrate that FundusGAN consistently outperforms state-of-the-art methods in terms of structural similarity, distribution alignment, and feature consistency. More importantly, our disease classification experiments confirm that augmenting training data with FundusGAN-generated images significantly improves diagnostic accuracy across multiple CNN architectures. These results establish FundusGAN as a valuable foundation model component that effectively addresses data scarcity challenges in ophthalmological AI research, enabling more robust diagnostic systems while reducing dependency on large-scale clinical data collection.

\begin{figure}[htbp]
    \centering
    \includegraphics[width=\textwidth]{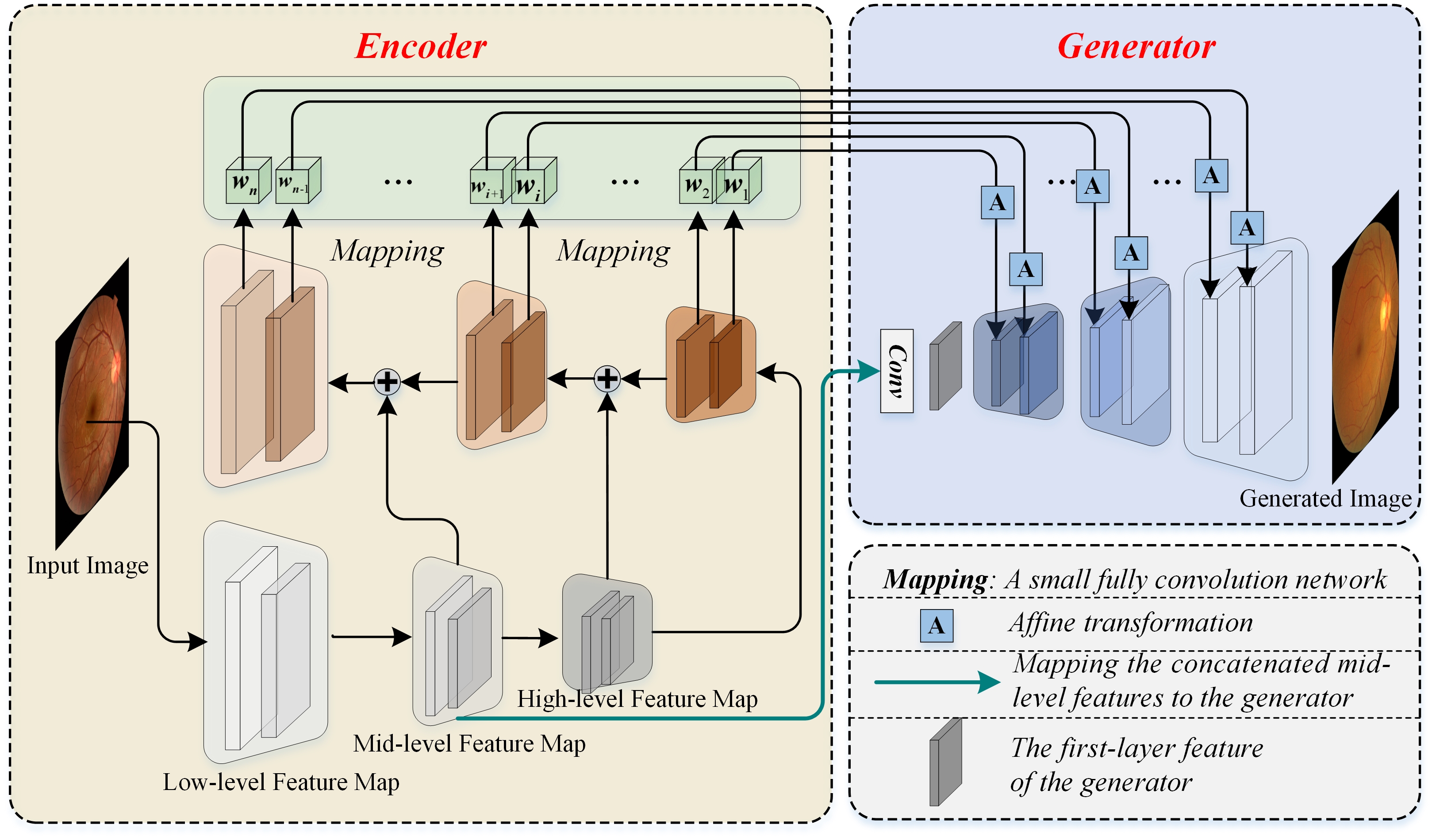}
    \caption{Architecture overview of fundusGAN. The framework comprises two core components: (1) A hierarchical encoder that processes input fundus images to extract multi-scale feature maps (low-, mid-, and high-level), and (2) A generator that generates output images through latent content vector space $w^+$ and first layer feature. }
    \label{fig1}
\end{figure}

\section{Method}
\subsection{The FundusGAN Framework}
\subsubsection{FundusGAN Encoder}

Given that fundus images contain critical structures at multiple scales (such as the optic disc, macula, vascular network, and micro-lesions), as shown in Figure~\ref{fig1}, FundusGAN incorporates a Feature Pyramid Network (FPN)~\cite{lin2017feature} in its encoder to comprehensively extract and fuse multi-scale information, thereby enhancing its capability to represent hierarchical features. Specifically, the encoder of FundusGAN first employs ResNet~\cite{he2016deep} as the backbone network to extract base features $F_b$, then utilizes the FPN to generate multi-scale feature maps. Through bottom-up feature extraction and top-down feature fusion, the FPN enables complementary interactions across different scales, significantly improving the model's perception of subtle lesions and complex fundus structures. In the encoder, low-level feature maps $F_{low}$ primarily capture low-frequency global information, such as structural characteristics of the optic disc and macula regions, which are typically large and exhibit stable anatomical features. Mid-level feature maps $F_{mid}$ focus on extracting morphological information of the vascular network, capturing the spatial distribution and branching patterns of major blood vessels. High-level feature maps $F_{high}$ specialize in fine-grained information extraction, particularly targeting microvasculature, hemorrhages, microaneurysms, and other small lesions, thereby enhancing the model's sensitivity to pathological changes. The related process can be formulated as:
\begin{equation}
    F_b = \text{ResNet}(x); \quad \{F_{low}, F_{mid}, F_{high}\} = \text{FPN}(F_b)
\end{equation}
where $x$ indicate the input fundus images. 

Based on the generated multi-scale feature maps $\{F_{low}, F_{mid},F_{high}\}$, we further train a compact mapping network $mapping(\cdot)$ to extract latent vectors $\{w_1,\cdots, w_n \}$ that encode distinct content information from $\{F_{low}, F_{mid},F_{high}\}$. The mapping network employs a lightweight fully convolutional network (FCN) to efficiently learn the mapping relationships between $\{F_{low}, F_{mid},F_{high}\}$ and $\{w_1,\cdots, w_n \}$. In the encoder, we define \( n = 18 \) latent vectors \( w^n \) to control the generation of diverse structural information in fundus images. Specifically, the mapping network extracts corresponding latent vectors from different levels of the feature pyramid to represent distinct structural features of fundus images: \\
• For high-frequency fine-grained structures such as microvasculature and small lesions, the $mapping(\cdot)$ generates latent vectors based on $F_{high}$ to capture subtle pathological variations. \\ 
• For the major vascular network, $F_{mid}$ are utilized to derive latent vectors, preserving morphological and branching details of blood vessels. \\ 
• For large-scale anatomical structures like the optic disc and macula, latent vectors are generated from $F_{low}$ to ensure stability and clarity in these critical regions. 

Through this design, FundusGAN adaptively decomposes and reconstructs hierarchical features of fundus images, achieving a balance between structural integrity and detail fidelity in generated images while ensuring diversity in fundus image generation. The related process can be formulated as:  
\begin{equation}
   w^+= \{w_1,\cdots,w_n\} = {mapping }(F_{low},F_{mid},F_{high})
\end{equation}

To ensure the generator effectively captures and reconstructs the primary structural information of fundus images, we choose to generate the base input features $f$ of generator $G(\cdot) $based on mid-level feature maps $F_{mid}$. These feature maps occupy a pivotal position in the feature pyramid, striking a balance between preserving global information of large-scale anatomical structures and encapsulating morphological and spatial distribution characteristics of the major vascular structure. This strategy provides the generator with stable and representative structural priors, enhancing the overall consistency of the generated images while maintaining anatomical plausibility and rich detail expression. Furthermore, due to mid-level feature maps inherit partial advantages from both low-level and high-level features, they effectively guide high-level features to refine the fine-grained representation of microvasculature and small lesions, while imposing structurally constrained priors on low-level features to suppress interference from irrelevant background noise. 
The related process can be formulated as:
\begin{equation}
    f = \text{Conv}(\text{Concat}(F_{mid})
\end{equation}

Additionally, to further optimize the detail generation in fundus images, we not only rely on the aforementioned latent vectors to govern the expression of structures at different scales but also introduce a hyperparameter \( \delta \) to modulate the number of skip connections between the encoder and the generator. By adjusting \( \delta \), we can flexibly control the weighting of multi-scale features during the generation process, enabling more precise reconstruction of anatomical structures and subtle lesions in fundus images. In the final generation pipeline, the foundational input features and the latent vectors encoding multi-scale characteristics are jointly fed into the generator. Each latent vector initiates its influence on the generation process through its corresponding affine transformation \( A_i \), adaptively adjusting feature distributions to ensure diversity and detail richness in the generated images. 
\begin{equation}
    \hat{x} = G(E_F(x,\delta), E_W(x)) := G(E(x, \delta))
\end{equation}
where $\hat{x}$ indicates generated fundus images, $E_F(\cdot)$ and $E_W(\cdot)$ denote the encoder extracting features and acquiring latent vectors, respectively. $G(\cdot)$ is the generator of FundusGAN.

\subsubsection{FundusGAN Generator} 
 
The generator $G(\cdot)$ of FundusGAN is designed based on StyleGAN~\cite{karras2019style} and optimized for the characteristics of medical fundus images to enhance the anatomical structure integrity and the representation of pathological features in the generated images.
Specifically, we removed upsampling operations starting from the 8th layer onward, ensuring that the feature resolutions of the first 7 layers remain consistent with the 7th layer. This adjustment aims to introduce more medically relevant structural information at lower-resolution stages, thereby stabilizing key anatomical regions such as the optic disc, macula, and vascular network.
However, conventional convolution kernels (or their receptive fields) at these layers may mismatch the scaled input feature resolutions, potentially leading to information loss or limited feature expressiveness. To address this, we replace these standard convolutions with dilated convolutions, which expand the receptive fields without increasing parameter count. For instance, in the first layer, we increase the dilation factor from 1 to 8, enabling the network to efficiently integrate large-scale information while maintaining computational efficiency.
Through this strategy, FundusGAN learns richer global contextual features at low-resolution stages and provides more robust feature representations for subsequent detail synthesis. This hierarchical optimization significantly improves the overall quality of generated fundus images and the precision of pathological details in lesion regions.

\subsection{Loss Functions}
 
The loss function of FundusGAN consists of two components to enhance the quality of generated fundus images while ensuring the anatomical consistency and authenticity of pathological features.
  
1) Regularization Loss  $\mathcal{L}_{\text{reg}}$

In the task of fundus image generation, the distribution of latent vectors directly influences the stability and consistency of the generated images. To prevent the generation of anomalous samples while ensuring the plausibility of key anatomical structures such as the optic disc, macula, and vascular network, we introduce a regularization loss. This loss encourages the encoder's output latent style vectors to be closer to a predefined average latent vector.  
This strategy is similar to the truncation trick~\cite{karras2019style} in StyleGAN, where the core idea is to constrain the distribution of latent vectors to generate clearer and more structurally stable fundus images without excessively restricting variability, which could otherwise compromise biological authenticity. 
In practice, this regularization loss is implemented by minimizing the Euclidean distance between the encoder’s output latent vector and the average latent vector \( \bar{w} \):
\begin{equation}
    \mathcal{L}_{\text{reg}} = \| E_W(x) - \bar{w} \|_2
\end{equation}
 
2) Reconstruction Loss  $\mathcal{L}_{rec}$

To ensure that the fundus images generated by FundusGAN are both realistic and consistent with medical conventions, we introduce a reconstruction loss to measure the pixel similarity between the generated image \( \hat{x} \) and the target real image \( x \). $\mathcal{L}_{rec}$ primarily consists of the following two components:

- Perceptual Loss $\mathcal{L}_{\text{LPIPS}}$: A pre-trained deep neural network (e.g., VGG) is used to extract features and measure the similarity between the generated image and the target image in a high-level feature space. This ensures the perceptual realism of the generated images. The corresponding formulation is as follows:
\begin{equation}
    \mathcal{L}_{\text{LPIPS}}(x) = \|Z(x) - Z(G(E(x)))\|_2
\end{equation}
where $Z(\cdot)$ is the perceptual feature extractor.

- Pixel-wise \( \mathcal{L}_2 \) Loss: This loss directly computes the mean squared error (MSE) between the generated image $\hat{x}$ and the target image $x$ in the pixel space, improving the accuracy of the overall structure. The corresponding formulation is as follows:
\begin{equation}
    \mathcal{L}_2(x) = \|x - G(E(x))\|_2
\end{equation}

Additionally, to further enhance the representation of fine structures in fundus images, such as microvasculature and small lesions, we optimize the latent vector \( w^+ \) and the generator’s first input feature \( f \).
\begin{equation}
 \hat{f}, \hat{w}^+ = \underset{f, w^+}{\operatorname{argmin}} \mathcal{L}_{\text{LPIPS}}(G(f, w^+), x)
\end{equation}
where $\hat{f}$ and $\hat{w}^+$ denotes the optimized latent vector and first input feature. 

This optimization strategy enables a more precise delineation of lesion areas and other high-frequency details while ensuring the overall anatomical rationality, thereby enhancing the medical value of the images generated by FundusGAN.
 
Finally, the total loss function $\mathcal{L}_{\text{total}}$ is defined as follows:
\begin{equation}
     \mathcal{L}_{\text{total}} = \lambda_1 \cdot \mathcal{L}_{\text{reg}} + \lambda_2 \cdot \mathcal{L}_{\text{LPIPS}}(x) + \lambda_3 \cdot \mathcal{L}_2(x)
\end{equation}

In this formulation, \( \lambda_1, \lambda_2, \lambda_3\) represent the weighting coefficients, and \( L_{\text{opt}}(w, f) \) represents the optimization term for the latent vector and input features.  

By integrating regularization loss and reconstruction loss, FundusGAN is capable of enhancing the clarity and detailed representation of lesion areas in the generated images while preserving the realistic anatomical structure of fundus images. This significantly improves its applicability in medical diagnosis and analysis.

\section{Experiments}
\subsection{Datasets and Implementation Details}

\subsubsection{DDR dataset~\cite{li2019diagnostic}:} The Diabetic Retinopathy Image Dataset (DDR) is a large-scale, clinically curated dataset designed for diabetic retinopathy (DR) severity grading. It comprises 13,673 high-quality fundus images collected from 147 hospitals across 23 provinces in China, ensuring broad geographic and demographic representation. Each image is classified into 5 severity levels based on the International Clinical Diabetic Retinopathy (ICDR) criteria: none, mild, moderate, severe, and proliferative DR. Low-quality images (originally categorized as a sixth class) are systematically excluded to enhance data reliability.\\

\subsubsection{ODIR dataset~\cite{li2021benchmark}:}The Ocular Disease Intelligent Recognition (ODIR) dataset is a multi-modal ophthalmic database containing 5,000 real-world patient records from multiple Chinese hospitals, including bilateral fundus images (left/right eyes), patient age, and diagnostic keywords. Collected using diverse commercial cameras (Canon, Zeiss, Kowa) with varying resolutions, it provides heterogeneous imaging conditions to reflect clinical practice. Each case is annotated by trained readers under quality control into 8 diagnostic categories: Normal (N), Diabetes (D), Glaucoma (G), Cataract (C), Age-related Macular Degeneration (A), Hypertension (H), Pathological Myopia (M), and Other abnormalities (O).

\subsubsection{DRIVE Dataset~\cite{staal2004ridge}:} DRIVE is a standard dataset for retinal vessel segmentation, containing 40 fundus images (resolution 584×565), divided into 20 images each for training and testing sets. Each image provides binary vessel masks annotated by two individuals, suitable for verifying the performance of vessel segmentation algorithms.

\subsubsection{IDRiD Dataset~\cite{porwal2018indian}:} IDRiD focuses on lesion localization and grading of diabetic retinopathy (DR), providing 81 high-resolution fundus images (4288×2848), including pixel-level annotations of four major lesions (microaneurysms, hemorrhages, hard exudates, soft exudates) and five-level DR severity classification, supporting multi-task model development.

\subsubsection{Implementation Details and Evaluation Metrics.}
FudnusGAN is implemented based on PyTorch and trained with 4 NVIDIA Quadro RTX 6000 GPUs. Considering the diversity and size of the input images, we resize the image size and patch size to 512 $\times$ 512 pixels. The Adam optimizer is applied to update parameters with a momentum of 0.9, an initial learning rate of 0.001, and a batch size of 8. A cosine annealing schedule is introduced to automatically adjust the learning rate. To comprehensively assess the performance of FundusGAN, we adopt three complementary metrics: 1) Structural Similarity Index (SSIM) quantifies the pixel-level structural consistency between generated and real fundus images, emphasizing the preservation of critical anatomical structures (e.g., optic disc, macula, and vascular topology). 2) Fréchet Inception Distance (FID) evaluates the global distributional alignment of synthesized images with real clinical data in the feature space, reflecting the model’s ability to generate clinically plausible and diverse pathological patterns. 3) Kernel Inception Distance (KID), a more robust variant of FID, measures feature distribution discrepancies without Gaussian assumptions, particularly sensitive to subtle pathological variations (e.g., microaneurysms, hemorrhages). These metrics collectively validate FundusGAN’s effectiveness in balancing anatomical fidelity (SSIM) and pathological authenticity (FID/KID), ensuring both structural coherence and diagnostically relevant feature synthesis.
\subsection{Comparison with the State-of-the-Art}

This section provides a quantitative comparison of fundus image generation methods on the DDR, DRIVE, and IDRiD datasets, demonstrating the effectiveness of the FundusGAN framework. We evaluate FundusGAN against several state-of-the-art methods, including Costa~\cite{costa2017end}, Guibas~\cite{guibas2017synthetic}, StyleGAN2~\cite{karras2020analyzing}, Diffusion~\cite{dhariwal2021diffusion}, Tub-sGAN~\cite{zhao2018synthesizing}, Patho-GAN 4×4~\cite{niu2021explainable}, Patho-GAN 3×3~\cite{niu2021explainable}, and MCML~\cite{yu2019retinal}. To ensure a fair and competitive evaluation, we strictly follow the experimental setups outlined in the original papers of each method.

\begin{table}[h]
\centering
\caption{Performance comparison of different models}
\label{tab:model_comparison}
\begin{tabular}{lccccc}
\toprule
\multirow{2}{*}{Models} & \multicolumn{3}{c}{DDR} & \multicolumn{1}{c}{DRIVE} & \multicolumn{1}{c}{IDRID}  \\
\cmidrule(lr){2-4} \cmidrule(lr){5-5} \cmidrule(lr){6-6}
 & SSIM & FID & KID & FID & FID \\
\midrule
Costa~\cite{costa2017end} & 0.6636 & 81.81 & 0.0918 & -- & -- \\
Guibas~\cite{guibas2017synthetic} & 0.6993 & 65.12 & 0.0713 & -- & -- \\
StyleGAN2~\cite{karras2020analyzing} & -- & -- & -- & 122.8 & -- \\
Diffusion~\cite{dhariwal2021diffusion} & -- & -- & -- & 86.78 & -- \\
Tub-sGAN~\cite{zhao2018synthesizing} & -- & -- & -- & -- & 117.82 \\
Patho-GAN4$\times$4~\cite{niu2021explainable} & -- & -- & -- & -- & 81.16 \\
Patho-GAN3$\times$3~\cite{niu2021explainable} & -- & -- & -- & -- & 80.13  \\
MCML~\cite{yu2019retinal} & -- & -- & -- & -- & --  \\
FundusGAN & 0.8863 & 54.2 & 0.0436 & 32.45 & 40.36 \\
\bottomrule
\end{tabular}
\end{table}

Table~\ref{tab:model_comparison} presents a quantitative comparison of various fundus image generation methods across the DDR, DRIVE, and IDRiD datasets. The evaluation metrics include Structural Similarity Index (SSIM), Fréchet Inception Distance (FID), and Kernel Inception Distance (KID), which collectively assess the visual quality, realism, and structural fidelity of the generated images.  
For the DDR dataset, FundusGAN achieves an SSIM of 0.8863, outperforming Costa (0.6636) and Guibas (0.6993), indicating superior structural preservation in generated images. Additionally, FundusGAN achieves a lower FID score (54.2) and KID score (0.0436) compared to Costa (FID: 81.81, KID: 0.0918) and Guibas (FID: 65.12, KID: 0.0713). These results suggest that FundusGAN generates images with higher perceptual quality and more realistic distributions, reducing the gap between synthetic and real fundus images.  
On the DRIVE dataset, FundusGAN achieves an FID score of 32.45, significantly outperforming StyleGAN2 (122.8) and Diffusion models (86.78). The large performance gap suggests that FundusGAN effectively captures and reconstructs vascular structures, which are critical for accurate fundus image synthesis. The ability to maintain fine-grained vascular patterns is crucial for applications such as segmentation and disease diagnosis, where vessel integrity plays a key role.  
For the IDRiD dataset, FundusGAN achieves an FID score of 40.36, performing better than Tub-sGAN (117.82) and approaching the results of Patho-GAN variants (81.16 and 80.13 for 4×4 and 3×3 configurations, respectively). The superior FID score of FundusGAN indicates that it generates fundus images with a distribution closer to real medical images, while its ability to maintain structural coherence is crucial for the accurate representation of retinal pathologies.  
Overall, the results demonstrate that FundusGAN consistently outperforms existing methods across multiple datasets, particularly in terms of structural preservation and visual realism. By incorporating multi-scale latent representations and feature extractions, FundusGAN effectively balances anatomical accuracy and perceptual quality, making it a promising solution for fundus image synthesis in medical applications.

\begin{figure}[htbp]
    \centering
    \includegraphics[width=0.8\textwidth]{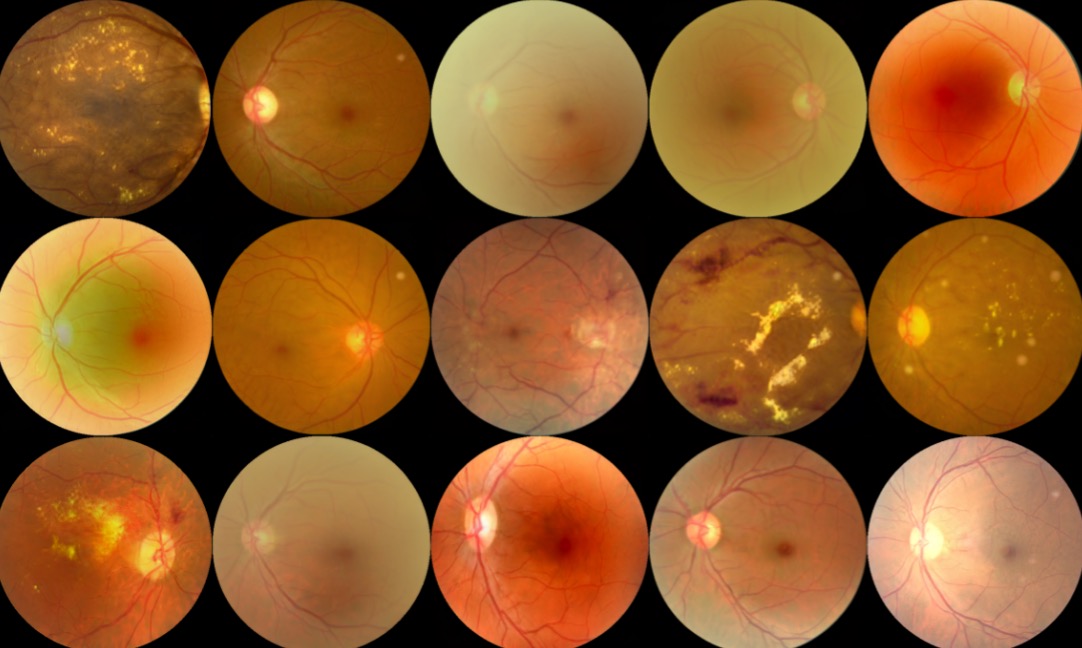}
    \caption{ Color fundus retinal images generated by FundusGAN. }
    \label{fig2}
\end{figure}

As shown in Figure~\ref{fig2}, we also show some cases of fundus images generated using FundusGAN. 
The images generated by FundusGAN exhibit outstanding anatomical fidelity. The branching patterns and spatial distribution of the retinal vascular network closely align with real clinical data, with precise simulation of both the curvature and connectivity of major and microvascular structures. The optic disc region maintains a well-defined circular boundary, and the macular area presents a natural color transition, without noticeable geometric distortions or anatomical misalignments. The color reproduction capability of the generated images is strong, with a distinct contrast between the orange-red background and the dark red blood vessels, consistent with the visual characteristics of real fundus images. Additionally, certain pathological features (e.g., hemorrhages and microaneurysms) are accurately simulated in both shape and location, resembling the typical manifestations of diabetic retinopathy.
Regarding multi-scale detail generation, the model demonstrates excellent local depiction capabilities. The textures of fine vascular termini are clearly visible, and the sharpness of microlesion boundaries is relatively high, ensuring that the overall image resolution meets medical diagnostic requirements. However, some generated images exhibit slight blurriness at the edges of microlesions, which may be attributed to insufficient high-level feature constraints. Additionally, the continuity of a few vascular branches requires further optimization. Future improvements could incorporate vascular topology priors to enhance structural coherence.
The overall layout and lighting conditions of the generated images effectively simulate the standardized requirements of real medical imaging, making them suitable for constructing large-scale fundus image databases or aiding in the training of algorithms for rare cases. Despite minor areas for improvement in fine details, FundusGAN achieves a high level of anatomical realism, pathological diversity, and visual consistency, providing high-quality data support for AI-assisted ophthalmic diagnostics.

\subsection{Ablation Study}

To investigate the contribution of each objective term to the performance of FundusGAN, we conduct a comprehensive ablation study by systematically removing each term on the DDR dataset. The results are reported in Table~\ref{tab:Ablation Study}, these results reveal the following several interesting points:

\begin{table}[h]
\centering
\caption{Ablation Study on the Contribution of Different Loss Components in FundusGAN}
\label{tab:Ablation Study}
\begin{tabular}{l|ccc|c|c}
\toprule
Method  & $\mathcal{L}_2$ & $\mathcal{L}_{\text{LPIPS}}$ & $\mathcal{L}_{\text{reg}}$ & FID & KID \\ 
\midrule
FundusGAN\_v1 & \checkmark & & & 72.6 & 0.0749 \\
FundusGAN\_v2 & \checkmark & \checkmark & & 60.3 & 0.0536 \\
FundusGAN & \checkmark & \checkmark & \checkmark & 54.2 & 0.0436 \\
\bottomrule
\end{tabular}
\end{table}

\textbf{1) Effect of Pixel-wise \(\mathcal{L}_2\)  Loss:} 
   FundusGAN\_v1, which only includes the pixel-wise \(\mathcal{L}_2\) loss, achieves the highest FID (72.6) and KID (0.0749), indicating that relying solely on pixel-level supervision leads to suboptimal image quality. This is expected, as \(\mathcal{L}_2\)  loss primarily enforces low-level similarity but lacks perceptual awareness, often resulting in overly smoothed images with limited fine detail.  

\textbf{2) Impact of Adding Perceptual Loss:} 

   Introducing the perceptual loss (\(\mathcal{L}_{\text{LPIPS}}\)) in FundusGAN\_v2 significantly improves the results, reducing FID to 60.3 and KID to 0.0536. This suggests that perceptual supervision enhances the realism of the generated images by preserving high-level structural and textural details, which are not captured by \(\mathcal{L}_2\)  loss alone.  

\textbf{3) Effect of Regularization Loss:}

   The full model (FundusGAN) incorporates the regularization loss (\(\mathcal{L}_{\text{reg}}\)) in addition to \(\mathcal{L}_2\) and perceptual loss, leading to the best performance with the lowest FID (54.2) and KID (0.0436). This demonstrates that constraining the latent space distribution helps generate more stable and realistic fundus images, likely by reducing artifacts and enforcing structural consistency in key anatomical regions.  

Overall, these results validate the importance of combining pixel-wise loss, perceptual loss, and regularization loss to achieve high-quality fundus image generation. The regularization term, in particular, plays a crucial role in further refining image fidelity and anatomical coherence.

\subsection{Multi-label Disease Classification Task Validation}
To evaluate the effectiveness of the generated images in disease grading, we conduct experiments by incorporating the generated images with real fundus images to train different models for disease classification.  
To assess the effectiveness of the generated images in disease grading, we conduct experiments by integrating synthetic images with real fundus images to train various deep learning models for disease classification.  
Specifically, we augment the original training dataset of ODIR by incorporating 5,000 fundus images generated by FundusGAN, forming an extended training set. We then train three widely used convolutional neural networks—DenseNet121, ResNet50, and VGG16—using this augmented dataset. The trained models are subsequently evaluated on the ODIR test set to analyze the impact of generated data on classification performance. The detailed results are presented in Table~\ref{tab:accuracy_comparison}.

\begin{table}[h]
\centering
\caption{Classification accuracy (\%) on different models using real and generated images.}
\label{tab:accuracy_comparison}
\begin{tabular}{lcccc}
\toprule
Accuracy (\%) & DenseNet121 & ResNet50 & VGG16 \\
\midrule
Real images  & 46.42 & 48.26 & 47.25 \\
Real + generated images & \textbf{51.26} & \textbf{54.75} & \textbf{52.28} \\
\bottomrule
\end{tabular}
\end{table}

Table~\ref{tab:accuracy_comparison} presents the classification accuracy of different models trained on real fundus images alone and on an extended dataset that includes both real and FundusGAN-generated images. The results demonstrate a consistent improvement across all tested models—DenseNet121, ResNet50, and VGG16—when trained with the augmented dataset.  
Specifically, DenseNet121 achieves an accuracy of 46.42\% when trained solely on real images, which increases to 51.26\% with the inclusion of generated images. Similarly, ResNet50 exhibits the most significant improvement, with accuracy rising from 48.26\% to 54.75\%. VGG16 also benefits from the synthetic data, achieving an increase from 47.25\% to 52.28\%.  
These findings suggest that the generated images effectively enrich the training data, enhancing the models' ability to generalize and recognize disease patterns more accurately.
Overall, the results validate the effectiveness of FundusGAN in generating high-quality fundus images that contribute to improved disease classification performance.
 
\section{Conclusion}
In this paper, we introduced FundusGAN, a novel hierarchical feature-aware generative framework designed specifically for high-fidelity fundus image synthesis. By incorporating a multi-scale feature extraction mechanism through Feature Pyramid Networks and optimized latent vector mapping, our approach effectively captures and reconstructs critical anatomical structures at various scales—from large-scale features like the optic disc to fine-grained details such as microvasculature and subtle lesions. The modified generator architecture with dilated convolutions and strategic adjustments to upsampling operations significantly enhances the anatomical fidelity and pathological representation in the generated images. Comprehensive experiments across multiple datasets (DDR, DRIVE, and IDRiD) demonstrate that FundusGAN consistently outperforms existing state-of-the-art methods in terms of structural similarity (SSIM), distribution alignment (FID), and feature consistency (KID). Furthermore, our ablation studies validate the importance of combining pixel-wise, perceptual, and regularization losses to achieve optimal generation quality. Most importantly, the practical utility of FundusGAN-generated images is confirmed through disease classification experiments, where models trained with augmented datasets show substantial performance improvements across different architectures. These results highlight FundusGAN's potential as a foundation model for addressing data scarcity challenges in ophthalmology, supporting both clinical training and AI-assisted diagnostic systems.

\bibliographystyle{splncs04}
\bibliography{refs-gen}
\end{document}